\documentclass[12pt]{iopart}
\pdfoutput=1
\usepackage{iopams}

\expandafter\let\csname equation*\endcsname\relax
\expandafter\let\csname endequation*\endcsname\relax

\usepackage{amsmath}
\usepackage{xcolor}
\usepackage{graphicx}
\usepackage{amssymb}

\newcommand{\ket}[1]{\vert #1 \rangle}
\newcommand{\eket}[1]{\bigl \vert #1 \bigr \rangle}

\newcommand{\ebra}[1]{\bigl \langle #1 \bigr \vert}

\newcommand{\figref}[1]{Fig.~\ref{#1}}

\newcommand{\ren}{R\'{e}nyi~}

\begin{document}

\title{Particle partition entanglement of one dimensional spinless fermions}

\author{Hatem Barghathi, Emanuel Casiano-Diaz and Adrian\\ {Del Maestro}}
\ead{Hatem.Barghathi@uvm.edu}
\address{Department of Physics, University of Vermont, Burlington, VT 05405, USA}

\begin{abstract}
We investigate the scaling of the R\'{e}nyi entanglement entropies for a
particle bipartition of interacting spinless fermions in one spatial dimension. In the Tomonaga-Luttinger liquid regime, we calculate the second R\'{e}nyi
entanglement entropy and show that the leading order finite-size scaling is
equal to a universal logarithm of the system size plus a non-universal
constant.  Higher-order corrections decay as power-laws in the system size
with exponents that depend only on the Luttinger parameter. We confirm the
universality of our results by investigating the one dimensional $t-V$ model of
interacting spinless fermions via exact-diagonalization techniques.  The
resulting sensitivity of the particle partition entanglement to boundary
conditions and statistics supports its utility as a probe of quantum liquids.
\end{abstract}

\maketitle

\section{Introduction}
Identical particles are fundamentally indistinguishable in quantum mechanics,
unlike their classical counterparts that can always be discriminated due to an
infinite set of observable properties. While this indistinguishability allows
for the power provided by the second quantization formalism, it can also lead to
ambiguity \cite{Zanardi:2001ec,Shi:2003jj,Shi:2004fw} when considering another
defining property of composite quantum systems: entanglement.  A pure state
representing $N$ quantum particles $\ket{\Psi} \in \mathcal{H}$ in Hilbert
space $\mathcal{H}$ is said to be bipartite entangled if it cannot be written
in a simple tensor product form $\ket{\Psi} \ne \ket{\Psi_{A}} \otimes
\ket{\Psi_{B}}$ where $A$ and $B$ are vector spaces with
$\ket{\Psi_{A}} \in A$ and $\ket{\Psi_B} \in B$ such that $A
\otimes B = \mathcal{H}$.  Conventionally, $A$ and 
$B$ correspond to a set of distinguishable single-particle modes whose
occupation numbers are physical observables, i.e.,  spatial or momentum modes.
However, for indistinguishable itinerant particles, there is no natural tensor
product decomposition into single-particle modes due to the
symmetrization or anti-symmetrization of the wavefunction with respect to the
interchange of first quantized particle coordinates for bosons and fermions,
respectively.  Thus, the \emph{mode entanglement}
may depend on the choice of single-particle modes, leading to questions as to which
(if any) are preferred and moreover, if these quantum correlations are
even physically meaningful \cite{Ghirardi:2004dl, Barnum:2004bm,
Dunningham:2005fu,Wiseman:2003vn, Wiseman:2003jx,
Benatti:2012cy,Balachandran:2013en, Dalton:2017qe}. For example, even in the absence of
interactions, a system of $N$ free itinerant bosons
\cite{Simon:2002it,Ding:2009gq} or fermions
\cite{Schliemann:2001ea,Zanardi:2002jw,Zanardi:2002gs} is always entangled
under a spatial biparition as a result of all allowed states being normalized
linear combinations of Slater determinants or permanents.  

Insights into these issues can be gained by considering the $N$-body
wavefunction in first quantized form  where a bipartition can be made in terms
of identical particle labels.  The resulting $n$-\emph{particle partition
entanglement} is a measure of quantum correlations between the subsets of $n$
and $N-n$ particles.  As individual (or groups of) identical
particles are not operationally distinguishable, there have been claims that
this type of entanglement is not useful as a resource for quantum information
processing \cite{Ghirardi:2004dl,Tichy:2011je,Balachandran:2013en}.  However,
schemes have been recently proposed to transfer it to experimentally
addressable modes \cite{Killoran:2014gu}.  In a foundational series of papers,
Haque \emph{et al.} explored the particle partition entanglement
in fractional quantum hall \cite{Zozulya:2007jw,Haque:2007il} and itinerant
bosonic, fermionic and anyonic lattice gases in one spatial dimension
\cite{Zozulya:2008kb,Haque:2009df}.  This type of particle partition
entanglement has since been investigated in other one dimensional systems
including the fermionic Calogero-Sutherland \cite{Katsura:2007hc}, anyonic
hard-core \cite{Santachiara:2007il} and bosonic Lieb-Liniger 
\cite{Herdman:2014jq, Herdman:2015gx} models
as well as rotating bose and fermi gases in two dimensions
\cite{Liu:2010pe}.  In analogy to the universal finite size scaling behavior of
the entanglement entropy of one dimensional quantum gases under a spatial
mode bipartition \cite{Calabrese:2004hl, Calabrese:2011fj,
Calabrese:2011ji}, a leading order scaling form for the particle partition
entanglement entropy $S$ supported by exact diagonalization on small lattice
models was proposed in Ref.~\cite{Zozulya:2008kb}  which is linear in the
subsystem size $n$ and logarithmic in the system size $N$: $S \sim n \ln N$.

Motivated by this empirical prediction, in this paper, we investigate the particle
partition entanglement for itinerant interacting spinless fermions in one
spatial dimension.  For Galilean invariant systems in the spatial continuum, we
confirm the scaling form proposed in Ref.~\cite{Zozulya:2008kb} within the
Tomonaga-Luttinger liquid framework \cite{Tomonaga1951, Haldane1981}
and determine how the leading order power-law corrections to the asymptotic
scaling depend on the strength of the interactions between particles for $n=1$.
By exploiting symmetries of the $n$-particle reduced density matrix, we are
able to measure the particle entanglement entropy in the one dimensional
fermionic $t-V$ model for systems composed of up to $M=28$ lattice sites at
half filling, allowing us to confirm our predictions from continuum field
theory. 

The rest of this paper is organized as follows. We introduce a quantitative
measure of entanglement, the \ren entanglement entropy and discuss some known
results for interacting spinless fermions.  We then derive the $1$-particle
entanglement entropy in the low energy limit and compare with exact
diagonalization results on a lattice.  We conclude with a discussion of the role
of boundary conditions, degeneracy and implications for future studies of
models with generalized statistics. All numerical data and code necessary to reproduce
the results and figures in this paper can be found in Ref.~\cite{repo}.

\section{Particle Partition Entanglement}
\begin{figure}[t]
\begin{center}
\includegraphics[width=0.7\columnwidth]{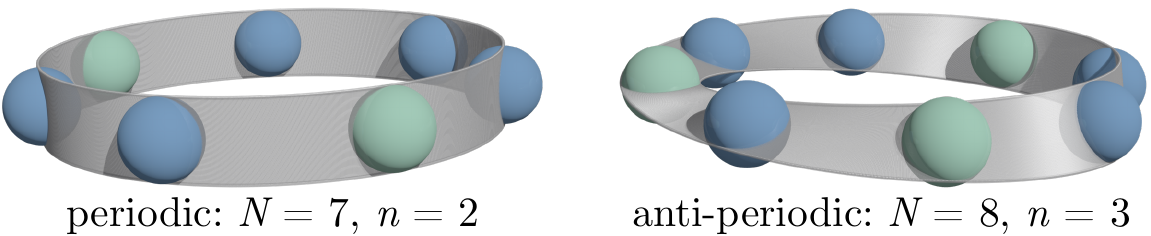}
\end{center}
\caption{A schematic of $N=7$ fermions in one spatial dimension subject to
periodic boundary conditions under a $n$-particle partition with $n=2$ 
(left) and anti-periodic boundary conditions with $N=8$ and $n=3$ (right).  All fermions
are identical, while the partitions $A$ and $B$ are distinguished via
their first quantized labels.}
\label{fig:particle_partition}
\end{figure}

The entanglement of the pure state $\eket{\Psi}$ under a general bipartition
into $A$ and $B$ can be quantified via the \ren entanglement entropy:
\begin{equation}
S_{\alpha}\left[ \rho_A \right] \equiv \frac{1}{1-\alpha} \ln \left({ {\rm Tr} 
\rho_A^{\alpha} }\right), 
\label{eq:renyi}
\end{equation}
where $\alpha$ is the \ren index and $\rho_A$ is the reduced density matrix
obtained by tracing out all degrees of freedom in $B$
\begin{equation}
    \rho_A \equiv {\Tr}_{B}\, \eket{\Psi}\ebra{\Psi}.
\end{equation}
For $\alpha = 1$ the \ren entropy is equivalent to the von Neumann
entropy: $-\Tr\, \rho_A \ln \rho_A$.   
%
%
While it is common for $A$ and $B$ to be defined by some set of observable modes, for a
many-body system consisting of $N$ itinerant particles they can refer to 
subsystems of particles.  As depicted in \figref{fig:particle_partition},
such a bipartition of indistinguishable particles (in this case spinless
fermions) is completely specified by the number of particles in the subsystem,
$n$. 
%
%
The entanglement entropy under a particle bipartition is then a function of the
familiar $n$-body reduced density matrix $\rho_n$, ($n$-RDM) defined in first
quantized notation in one spatial dimension as: 
\begin{equation}
\rho_n \equiv \int dx_{n+1} \cdots \int d x_{N}\left
\langle {x}_{n+1} \cdots {x}_{N} \vert \Psi \right
\rangle \left \langle \Psi \vert {x}_{n+1}\cdots {x_{N}} \right \rangle 
\label{eq:rhon}
\end{equation}
where we have taken the normalization ${\rm Tr} \rho_n = 1$. From this form, it
is clear that the particle partition \ren entropies $S_\alpha[\rho_n] \equiv
S_\alpha(n)$ only vanish when the $N$-body ground state $\eket{\Psi}$ can be
written as a general tensor product state in first quantized notation. This
immediately implies that $S_\alpha(n)=0$ when all particles are condensed into
a single mode, and thus the particle partition entanglement of
the non-interacting Bose gas is identically zero, in contrast to non-zero
results for its spatial mode entanglement \cite{Simon:2002it, Ding:2009gq}.
This is not the case for many-fermion systems, which always have non-zero
particle entanglement, even in the absence of interactions \cite{Zanardi:2002jw}.  Particle
entanglement entropy is sensitive to both interactions and 
statistics, and as $\rho_n$ is free of any length scale, it can capture
non-local effects making it complimentary to the more conventionally studied
spatial mode entanglement entropy.

As described in the introduction, Zozulya et al.  \cite{Zozulya:2008kb} first
proposed a ``standard'' finite-size scaling form for the particle entanglement
entropy of fermions:
\begin{equation}
    S(n,N) =  \ln \binom{N}{n} + a + \mathcal{O}\left(\frac{1}{N^\gamma}\right)
\label{eq:SNScaling}
\end{equation}
where $a$ and $\gamma$ are non-universal dimensionless numbers that can depend on $n$.  These
coefficients are known for the case of non-interacting fermions where $a=0$
\cite{Haque:2009df} and for the Laughlin state with filling fraction $\nu$: $a
= -n \ln \nu$, $\gamma=2$ when $n \ll N$ \cite{Zozulya:2007jw}.  

Recently, a general scaling form like Eq.~(\ref{eq:SNScaling}) was investigated
for a system of interacting bosons in the spatial continuum with $n=1$
\cite{Herdman:2015gx} where it was found that the pre-factor of the leading
order logarithm is non-universal, depending on the interaction
strength.  In this paper, we apply extensions of these methods to
interacting Galilean invariant one dimensional fermions and are able to 
systematically derive Eq.~(\ref{eq:SNScaling}) while presenting results for
both $a$ and $\gamma$ as a function of the interaction strength.

\section{One-particle entanglement in fermionic Tomonaga-Luttinger liquids}
We are interested in the asymptotic finite size scaling of the entanglement
entropy (EE) as defined in Eq.~(\ref{eq:renyi}) which can be investigated for
any \ren index $\alpha$.  Here we focus on the special case of $\alpha=2$ as
(i) the calculation will turn out to be analytically tractable and
(ii) as it can be related to the expectation value of a local observable, it
has proved to be the most direct numerical
\cite{Hastings:2010dc,Grover:2013cs,McMinis:2013dp,Drut:2015fs} and even experimental
\cite{Islam:2015cm,Melko:2016bo} route to its measurement. We begin by considering a system
of $N$ one-dimensional interacting spinless fermions with density $\rho_0=N/L$ (where $L$
is the length of the system) whose low energy properties can be described in
terms of the universal quantum hydrodynamics of 
Tomonaga-Luttinger liquid (TLL) theory \cite{Tomonaga1951, Haldane1981}.
Within this framework, at zero temperature in the thermodynamic limit, any
$n$-body reduced density matrix can in principle be computed
\cite{giamarchi:2004qu} and in particular for $n=1$ \cite{Dzyaloshinskii:1974un}
\begin{equation}
\rho_1\left(x,x'\right)=\frac{\sin(\pi \rho_0|x-x'|)}{\pi
\rho_0L|x-x'|(1+|x-x'|^2\Lambda^2)^{(K+K^{-1}-2)/4}},
\label{eq:TLLOBDM}
\end{equation}
where $\Tr \rho_1 = 1$ and both the ultraviolet (inverse short-distance) cutoff
$\Lambda$ and TLL parameter $K$ depend on the microscopic details of the
interaction between particles. Specifically, $K$ characterizes the nature of
the interaction, where $0<K<1$ ($K>1$) corresponds to repulsive (attractive)
interactions with free fermions having $K=1$.  For ease of notation, we will
replace the non-negative $K$-dependent exponent in Eq.~(\ref{eq:TLLOBDM}) with
$g\equiv(K+K^{-1}-2)/4$.  

The one-particle partition second \ren entanglement entropy can be computed by using
$\rho_1$ in Eq.~(\ref{eq:renyi}) 
\begin{eqnarray}
S_2(n = 1)=-{\rm{ln}}\left({\rm{Tr}}\left[{\rho}^2_1\right]\right)&=&
    -{\rm{ln}}\left(\int_{-L/2}^{L/2} dx \int_{-L/2}^{L/2}dx'
    \rho_1\left(x,x'\right)\rho_1\left(x',x\right)\right)\nonumber\\ &=&
    {\rm{ln}}(N)-{\rm{ln}}(f(N,g,\Lambda/\rho_0)),\label{eq:S2_TLL}
\end{eqnarray}
where we have used translational invariance of the system and 
\begin{align}
    f(N,g,\Lambda/\rho_0) & =\int_{0}^{\infty} dy\frac{2\sin^2(\pi
y)}{\pi^2y^2(1+y^2\Lambda^2/\rho_0^2)^{2g}} \nonumber \\
& \quad -\int_{N/2}^{\infty} dy\frac{2\sin^2(\pi y)}{\pi^2y^2(1+y^2\Lambda^2/\rho_0^2)^{2g}}. 
\label{eq:f_TLL_1}
\end{align}
The first integral can be evaluated exactly in terms of special functions:
\begin{align}
A (g,\Lambda/\rho_0) &= \int_{0}^{\infty} dy\frac{2\sin^2(\pi
y)}{\pi^2y^2(1+y^2\Lambda^2/\rho_0^2)^{2g}} \nonumber \\
&= \frac{ \pi ^{4 g+\frac{1}{2}} \rho_0 ^{4g} \sec (2 \pi
g) \, _1F_2\left(2 g;2 g+1,2 g+\frac{3}{2};\pi ^2 \Lambda ^{-2}\rho_0
^2\right)}{2 \Lambda ^{4 g} \Gamma (2 g+1) \Gamma (2 g+\frac{3}{2})}\nonumber\\
&\quad\quad + \frac{ \Lambda\Gamma \left(2 g+\frac{1}{2}\right) \left[\,
_1F_2\left(-\frac{1}{2};\frac{1}{2},\frac{1}{2}-2 g;\pi ^2 \Lambda ^{-2} \rho_0
^2\right)-1\right]}{\pi ^{3/2} \rho_0 \Gamma (2 g)}.
\label{eq:A_TLL}
\end{align}
where $_1F_2(q;c,d;z)$ is the generalized hypergeometric and $\Gamma(z)$ the
Gamma function.  The leading order $N$ dependence of the
second integral in Eq.~(\ref{eq:f_TLL_1}) can be extracted by replacing the
highly oscillating periodic function $\sin^2(\pi y)$, in the large $N$ limit,
by its average over one period, i.e., $\sin^2(\pi y) \approx 1/2$ and
expanding the rest of the integrand for large $y$. We find
\begin{equation}
f(N,g,\Lambda/\rho_0) \simeq  A(g,\Lambda/\rho_0)-\frac{2^{4 g+1}}{\pi ^2 (4
g+1)(\Lambda/\rho_0)^{4g}} \frac{1}{N^{4g+1}} 
\label{eq:f_infty}
\end{equation}
and thus the second \ren EE for $n=1$ has the asymptotic form
\begin{equation}
S_2(n=1) =  \ln(N)-\ln\left[A(g,\Lambda/\rho_0)\right] +
\frac{b(g,\Lambda/\rho_0)}{N^{4g+1}}
+\mathcal{O}\left(\frac{1}{N^{4g+2}}\right)
\label{eq:s2n1_asymptotic}
\end{equation}
where
\begin{equation}
b(g,\Lambda/\rho_0)=\frac{2^{4 g+1}}{\pi ^2 (4 g+1)(\Lambda/\rho_0)^{4g}
A(g,\Lambda/\rho_0)}.
\label{eq:b_TLL} 
\end{equation}
This result constitutes an analytical confirmation of the empirical scaling
form in Eq.~(\ref{eq:SNScaling}) first proposed by Haque \emph{et al.}
\cite{Zozulya:2008kb,Haque:2009df}, with $n=1$, where  
\begin{equation}
a=-\ln\left[A(g,\Lambda/\rho_0)\right],\; \gamma=4g+1.
\label{eq:TLL_a1_gamma}
\end{equation}

\subsection{Non-interacting spinless fermions}

In the non-interacting limit when $K=1$ ($g=0$), Eq.~(\ref{eq:A_TLL}) yields
$A(0,\Lambda/\rho_0)=1$ and thus $a=0$ in agreement with previous calculations
of the particle partition EE for free fermions (FF) on a lattice
\cite{Zozulya:2008kb} where it was found that $S_{2,{FF}}(n=1) = \ln N$.
However, combining Eqs.~(\ref{eq:s2n1_asymptotic})-(\ref{eq:b_TLL}) for $g=0$
yields 
\begin{equation}
S_{2}(n=1)\simeq\ln(N)+\frac{2}{\pi^2N}.
\label{eq:s2n1_FF_asymptotic}
\end{equation}
in disagreement with the lattice result by a factor of $\mathcal{O}(N^{-1})$.  To
ensure that this discrepancy does not arise from the approximations made in
expanding the integral in Eq.~(\ref{eq:f_TLL_1}) we can return to the exact expression for the
1-RDM for non-interacting spinless fermions:
\begin{equation}
\rho_{1,FF}\left(x,x'\right)=\frac{\sin(\pi \rho_0|x-x'|)}{\pi \rho_0L|x-x'|},\label{eq:FFOBDM}
\end{equation}
which leads to a soluble integral and analytic form for the EE in the spatial
continuum:
\begin{equation}
    S_{2,FF}(n=1)=\ln(N)-\ln\left\{\frac{2 \left[N \pi {\rm{Si}}(N \pi )+\cos (\pi
    N)-1\right]}{\pi ^2 N}\right\}
\label{eq:s2n1_FF}
\end{equation}
where $\mathrm{Si}(z)$ is the sine integral.  Expanding for large $N$ recovers
the asymptotic form in Eq.~(\ref{eq:s2n1_FF_asymptotic}) which differs from the
known lattice result.

\subsection{Effects of boundary conditions}

The origin of this $1/N$ difference between free spinless fermions in the
continuum vs. the lattice is related to our neglect of finite-size boundary
conditions when studying the asymptotic behavior of the second \ren EE.  To
properly capture the finite-size effects of periodic boundary conditions we
replace separations $\lvert x-x'\rvert$ with the chord length between two
points on a ring of circumference $L$ \cite{Cazalilla:2004}:
\begin{equation}
|x-x'| \rightarrow \frac{L}{\pi}\sin\left(\frac{\pi} {L}|x-x'|\right).
\end{equation}
Using the finite-size corrected 1-RDM, the integral in Eq.~(\ref{eq:f_TLL_1}) takes the form
%
\begin{equation}
    f(N,g,\Lambda/\rho_0)= \frac{2}{N^2}\int_{0}^{N/2} dy\frac{\sin^2(\pi
y)}{\sin^2(\frac{\pi
y}{N})\left[1+\frac{N^2\Lambda^2}{\pi^2\rho_0^2}\sin^2(\frac{\pi y}{N})\right]^{2g}}. 
\label{eq:f_TLL_3}
\end{equation}
where the effects of finite $L$ will appear only in the prefactors of 
decaying terms in an asymptotic expansion.  Employing Eq.~(\ref{eq:f_TLL_3})
for free fermions with $g=0$ we recover the known lattice result 
$S_{2,FF}(n=1)=\ln(N)$.  For all subsequent comparisons with numerical data at
finite $g$ we employ the appropriately finite size corrected form of the 1-RDM
when computing the \ren entanglement entropy.
%

\section{Exact diagonalization of the $t-V$ chain of spinless fermions}
In order to test the validity of our main result in
Eq.~(\ref{eq:s2n1_asymptotic}) for the $n=1$ particle partition EE, we consider
the $t-V$ model of $N$ spinless fermions on a chain with $M$ sites defined by
the Hamiltonian
\begin{equation}
  H= -t\sum_{i}\left(c^\dagger_{i} c^{\phantom{\dagger}}_{i+1} +c^\dagger_{i+1} c^{\phantom{\dagger}}_{i} \right) +V\sum_i n_i n_{i+1}\,
  \label{eq:H-tV}
\end{equation}
where $c^\dagger_{i}$ and $c^{\phantom{\dagger}}_{i}$ are the fermionic
creation and annihilation operators at site $i$ and
$n_i=c^\dagger_{i}c^{\phantom{\dagger}}_{i}$ is the occupation number.
The model is parameterized by the nearest-neighbor hopping amplitude $t>0$, and
interaction strength $V$. We consider only the half-filled case ($M=2N$) with
periodic boundary conditions (PBC) for odd number of fermions $N$, while for
even $N$ we use antiperiodic boundary conditions (APBC) 
to avoid the otherwise degenerate ground state
\cite{Cazalilla:2004} (See Fig.~\ref{fig:particle_partition}). In order to make
connection with the general TLL theory described above, we require a method to
determine the parameter $K$ from the microscopic $t-V$ model.  This can be
accomplished via the Jordan--Wigner transformation
\cite{Jordan:1928} which maps the $t-V$ model onto the XXZ spin-1/2
chain that is exactly solvable \cite{DesCloizeaux:1966,DesCloizeauxGaudin:1966}. In the range
$|V/t|<2$, the system is known to be in the TLL phase, where the analytical form of $K$
is given by
\begin{equation}
    K=\frac{\pi}{2\cos^{-1}(-V/2t)}.
\label{eq:K_tV}
\end{equation}
By increasing the repulsive interaction across $V/t=2$ ($K=1/2$), the system
undergoes a continuous phase transition to a charge-density wave (CDW) phase.
In contrast, the transition across $V/t=-2$ ($K\rightarrow \infty$) is a
discrete one, where the fermions tend to form a single cluster.

\begin{figure}[h]
\begin{center}
\includegraphics[width=0.7\columnwidth]{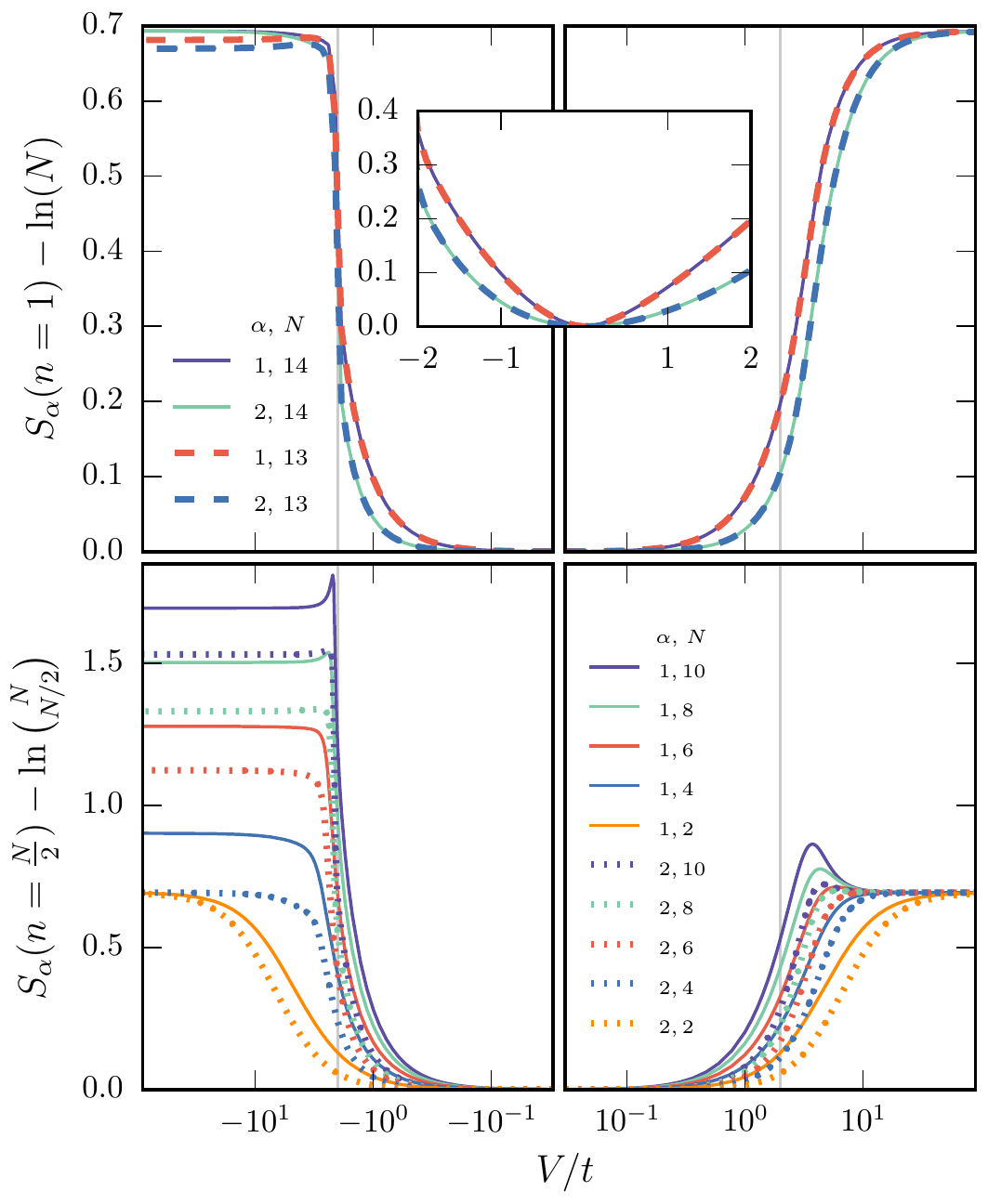}
\llap{\parbox[b]{0.4in}{\textbf{(a)}\\\rule{0ex}{4.85in}}}
\llap{\parbox[b]{0.45in}{\textbf{(b)}\\\rule{0ex}{2.55in}}}
\end{center}
\caption{Interaction effects on the $n$-particle entanglement entropy
$S_\alpha(n)$ for $\alpha=1,2$ in the ground state of the $t-V$ model. (a)
$S_{\alpha}(n=1)-\ln N$ vs $V/t$ for $N=13$ and $14$ with periodic and
anti-periodic boundary conditions, respectively. The light gray vertical lines
mark the location of the known phase transitions at $V/t=\pm 2$. The
subtracted $\ln (N)$ term is the one-particle entanglement
entropy for free fermions.  Inset: the Tomonaga-Luttinger liquid
region where we expect the continuum theory to apply. (b)
$S_{\alpha}(n=N/2)-\ln \binom{N}{N/2}$ vs
$V/t$ for \emph{macroscpic} partitions with $n = N/2$ and anti-periodic
boundary conditions. As $N$ increases, features appear near the phase transitions
for $\alpha=1$.} 
\label{fig:EntropiesFigure} 
\end{figure}

Beginning with the non-interacting case ($V/t=0$), the FF 
Hamiltonian is diagonal in the momentum-space representation leading to a
ground state that is a Slater determinant of the $N$ lowest energy modes. The
rank of the resulting $n$-RDM is $\binom{N}{n}$ and with equal eigenvalues
\cite{Zozulya:2008kb}, it follows (as introduced above) that all the \ren EEs are equal to   
\begin{equation}
S_{\alpha,FF}(n) =\ln{\binom{N}{n}}.
\label{S_FF_tV} 
\end{equation}
In the presence of interactions, we calculate the von Neumann $(\alpha=1)$ and
the second $(\alpha=2)$ \ren EEs from the ground state of Eq.~(\ref{eq:H-tV})
which we obtain via numerical exact diagonalization. The resulting $n$-RDM has
maximum possible rank $\binom{M}{n}$ due to the indistinguishability of the
$n<N$ particles in the partition, as opposed to $n!\binom{M}{n}$, the full
dimension of the Hilbert space in the first quantized basis.  Exploiting this
symmetry, (for details, see \ref{appendixA}) we are able to study systems up to
$M=28$ sites, a considerable advancement over previous work
\cite{Haque:2009df}.  The results are shown in \figref{fig:EntropiesFigure}
%
%
which demonstrates that the entanglement entropy $S_{\alpha}(n=1)$ increases
with increasing interaction strength $|V/t|$ up to a maximum of $
S_{\alpha,FF}(n=1)+\ln 2$ (for even $N$) in the limit $|V/t| \rightarrow
\infty$ \cite{Zozulya:2008kb,Haque:2009df}.  For attractive interactions,
$S_{\alpha}(n=1)$ displays a sharp increase around the first-order transition
point $V/t = -2$. In contrast, $S_{\alpha}(n=1)$ does not seem to be sensitive
to the continuous transition at $V/t=2$ \cite{Zozulya:2008kb}. However,
when considering a macroscopic partition size $n=N/2$, we observe that
$S_{\alpha}(n=N/2)$ develops a peak near $V/t=2$ which appears to
approach the critical point as we increase $N$ (\figref{fig:EntropiesFigure}
(b)). Eventually, $S_{\alpha}(n=N/2)$ saturates to $\ln \binom{N}{N/2}+\ln2$ in
the limit $V/t\rightarrow\infty$, with details given in \ref{appendixB}. 
 
\begin{figure}[h]
\begin{center}
\includegraphics[width=0.7\columnwidth]{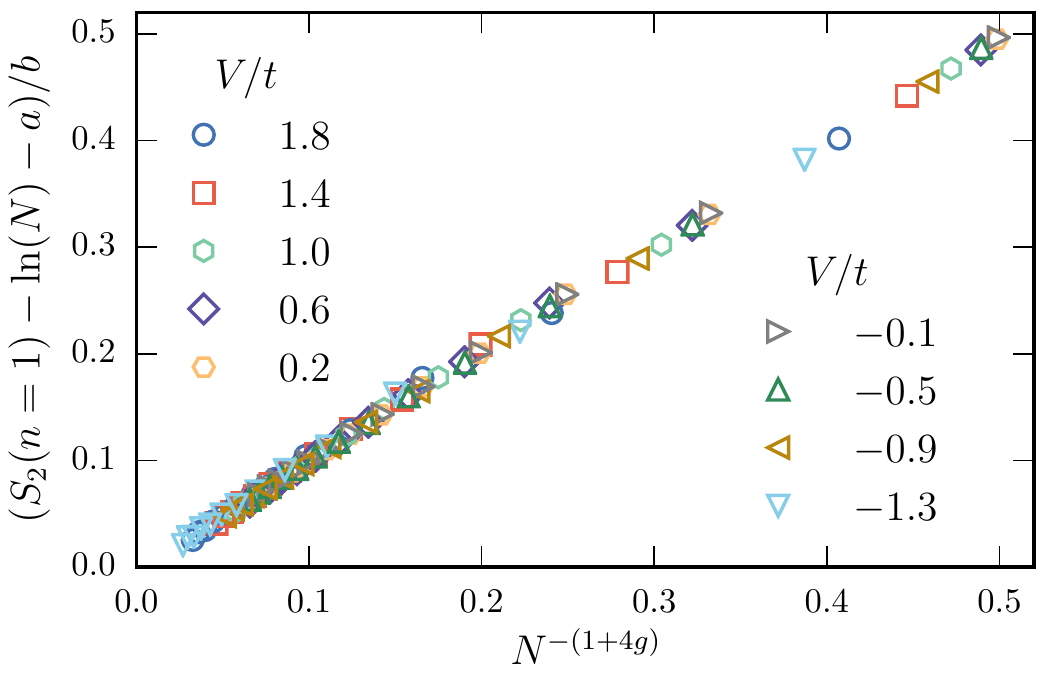}
\end{center}
\caption{Finite size scaling of $S_2(n=1)-\ln(N)$ with $N^{-(4g+1)}$
for $2 \leq N \leq 14$ confirming the empirical asymptotic scaling predicted by
Zozulya \emph{et al.} \cite{Zozulya:2007jw} and identifying the power of the
leading finite size correction as $\gamma = 4g + 1$.  The coefficients $a$ and
$b$ depend on the interaction strength $V/t$ and are calculated from a linear
fit of the exact diagonalization data according to Eq.~(\ref{eq:s2n1_asymptotic}).} 
\label{fig:Scaling_of_PPE_In_TLL_phase}
\end{figure}

We now turn to the  TLL region $|V/t| < 2$, where we expect the scaling of the interaction
contribution to the EE: $S_2(n=1)-\ln(N)$, to be linear in $1/N^{4g+1}$ with
corrections of $\mathcal{O}(1/N^{4g+2})$ as in Eq.~(\ref{eq:s2n1_asymptotic}).
To test this prediction, we rearrange Eq.~(\ref{eq:s2n1_asymptotic}) as:
\begin{equation}
\frac{S_2(n=1)-\ln(N)-a}{b}=N^{-(4g+1)}+\mathcal{O}\left(N^{-(4g+2)}\right).\label{eq:s2n1_scaling_tv}
\end{equation}
and calculate $S_2(n=1)$ as a function of $N$ using the ground state of $t-V$
model for different values of the interaction strength $V/t$, deep in the TLL
phase (away from the phase transitions). For each interaction strength $V/t$, we
compute $g=(K+K^{-1}-2)/4$ using Eq.~(\ref{eq:K_tV}) and extract $a$ and $b$
from a linear fit to the $S_2(n=1)-\ln(N)$ vs $ N^{-(4g+1)}$ data set.
Next, we use the extracted coefficients to rescale $S_2(n=1)-\ln(N)$ according
to Eq.~(\ref{eq:s2n1_scaling_tv}). The results are illustrated in
\figref{fig:Scaling_of_PPE_In_TLL_phase}, where, for suitably large $N$, the data follows the straight line
predicted by Eq.~(\ref{eq:s2n1_scaling_tv}) with unit slope, verifying the TLL
scaling form in Eq.~(\ref{eq:s2n1_asymptotic}). Deviations from linearity 
for smaller $N$ arise due to finite size corrections of $\mathcal{O}(1/N^{4g+2})$.
%
%

\begin{figure}[h]
\begin{center}
\includegraphics[width=0.7\columnwidth]{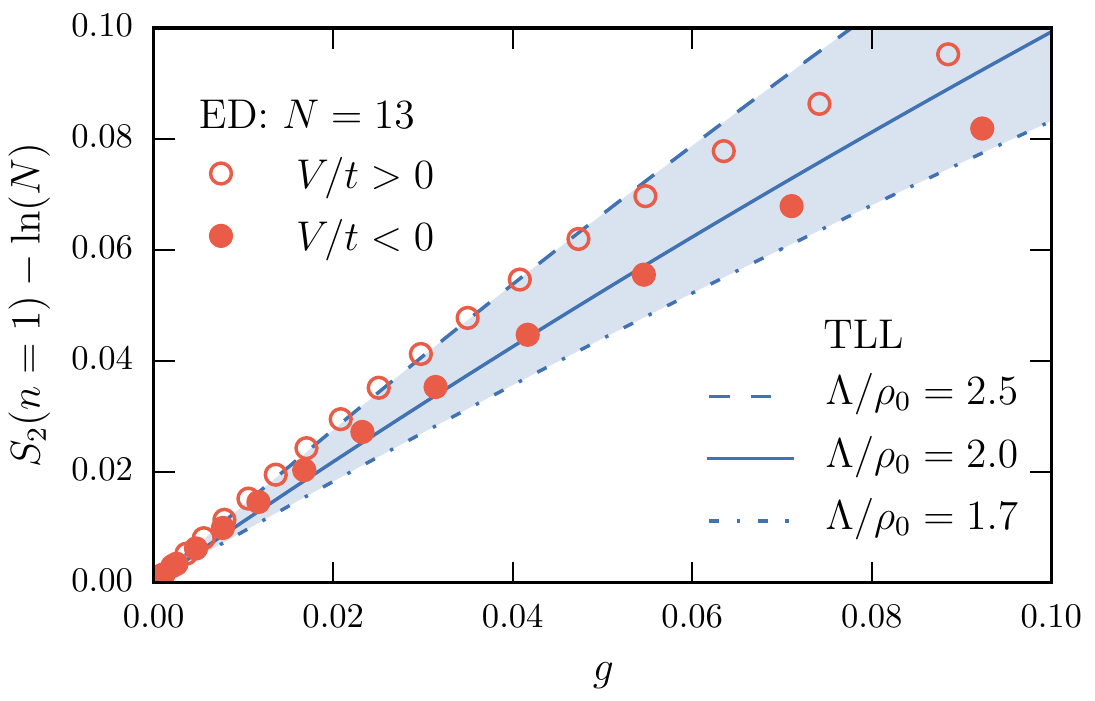}
\end{center}
\caption{The effective interaction dependence of the $1$-particle partition
second \ren entanglement entropy $S_2(n=1)-\ln(N)$.  Open (closed)
points were computed via exact diagonalization of the $t-V$ model for $N=13$
with repulsive (attractive) interactions.  The lines show the prediction from
the Tomonaga-Luttinger liquid theory for three different values of the
ultraviolet cutoff $\Lambda$ measured in units of the density $\rho_0$.} 
\label{fig:Svsg}
 \end{figure}

Having understood the asymptotic scaling of the 1-particle partition \ren EE
with $N$, we now consider its dependence on the interaction strength $g$.  This
amounts to asking if the $g$-dependence of the scaling coefficients $a$ and $b$
for the $t-V$ model can be predicted from our continuum theory.  To answer this
question we calculate the second \ren EE for $|V/t| < 2$ in the liquid phase at
fixed $N$ by evaluating the full integral in Eq.~(\ref{eq:f_TLL_3}) numerically
including all contributions from finite $N$.
However, in order to compare the resulting particle EE with that obtained from
the exact diagonalization, we need to identify the corresponding non-universal
value of the ratio $\Lambda/\rho_0$ in the $t-V$ model. At half filling, the
average particle density is $\rho_0=1/2x_0$ where $x_0$ is the lattice
separation, while one estimates the ultraviolet cutoff $\Lambda$ to be of the
order of $1/x_0$, yielding $\Lambda/\rho_0 \approx 2$.  The open and closed
symbols in Fig.~\ref{fig:Svsg} show the exact diagonalization results for
$S_2(n=1)-\ln(N)$ as a function of $g$ for $N=13$.  
%
%
The three lines correspond to the prediction from the TLL theory for different values of the UV cutoff $\Lambda$. Due to the highly non-linear
relationship between the interaction strength $V/t$ and the TLL parameter $K$
(Eq.~\ref{eq:K_tV}), in combination with the sensitivity of the particle
partition entanglement to the strength and nature of inter-particle
interactions, it is no surprise that the EE in the $t-V$ model is a
multi-valued function of the effective interaction parameter $g$ for attractive
and repulsive interactions. Clearly, high energy lattice-scale physics, not
captured within the low energy TLL theory is responsible for this behavior.
Moreover, recall that the ultraviolet cutoff, $\Lambda$, in
Eq.~(\ref{eq:TLLOBDM}), is proportional to the inverse of the effective range
of the interaction \cite{Dzyaloshinskii:1974un}.  Therefore, we expect
$\Lambda$ to exhibit a dependence on the nature and strength of the
interaction, i.e., have $K$-dependence  \cite{Herdman:2015gx}.  Considering
such a dependence, we find that the $t-V$ model results for $S_2(n=1)-\ln(N)$
are bounded by the theoretically calculated ones using $\Lambda/\rho_0=1.7$ and
$2.5$ (Fig.~\ref{fig:Svsg}). Note that both ratios are of order $2$.

\begin{figure}[h]
\begin{center}
\includegraphics[width=0.7\columnwidth]{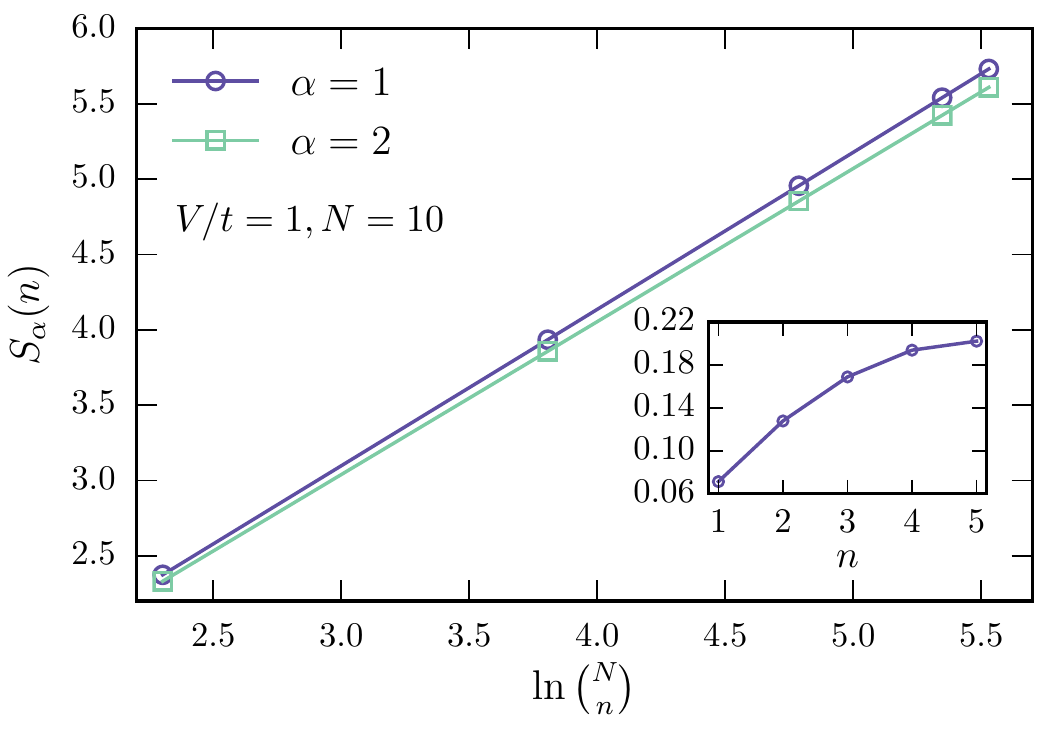}
\end{center}
\caption{Scaling of $S_\alpha(n)$ with $\ln \binom{N}{n}$ for $\alpha=1,2$ in
the ground state of the $t-V$ model with $V/t=1$, $N=10$, and for partition
sizes $1 \leq n \leq 5$.  Inset: Interaction  contribution to the EE
($S_1(n)-\ln \binom{N}{n}$) vs $n$.} 
\label{fig:SvsNchoosen}
 \end{figure}

Testing the proposed leading order scaling of the particle partition
EE in Eq.~(\ref{eq:SNScaling}) with the partition size $n$ in the TLL phase,
requires the calculation of the $n$-RDM with $n>1$.  While this can be done in
principle using standard techniques \cite{giamarchi:2004qu}, the resulting
evaluation of $S_2(n)$ requires performing $2n$ non-separable integrals.  Even
for the $n=2$ case we were not able to analytically extract the asymptotic scaling
of $\rm{Tr}\ \rho_2^2$.  However, from numerical exact diagonalization of the
$t-V$ model in the in the TLL phase we were able to calculate
the \ren EEs for partitions up to $n=N/2=5$ for $N=10$ 
as seen in Fig.~\ref{fig:SvsNchoosen}. 
%
%
Our results are in agreement with previous calculations of $N=6,n=3$
\cite{Zozulya:2008kb} and strongly suggest that the leading term in the scaling
of the \ren EEs with $n$ is indeed equal to the \ren EE of free fermions, i.e.,
$\ln \binom{N}{n}$.  Interactions introduce a correction term that increases
with the partition size with a negative curvature (see Fig.~\ref{fig:SvsNchoosen}
inset) such that both the leading order constant and finite-size power-law
corrections to scaling both depend on $n$.

\begin{figure}[h]
\begin{center}
\includegraphics[width=0.7\columnwidth]{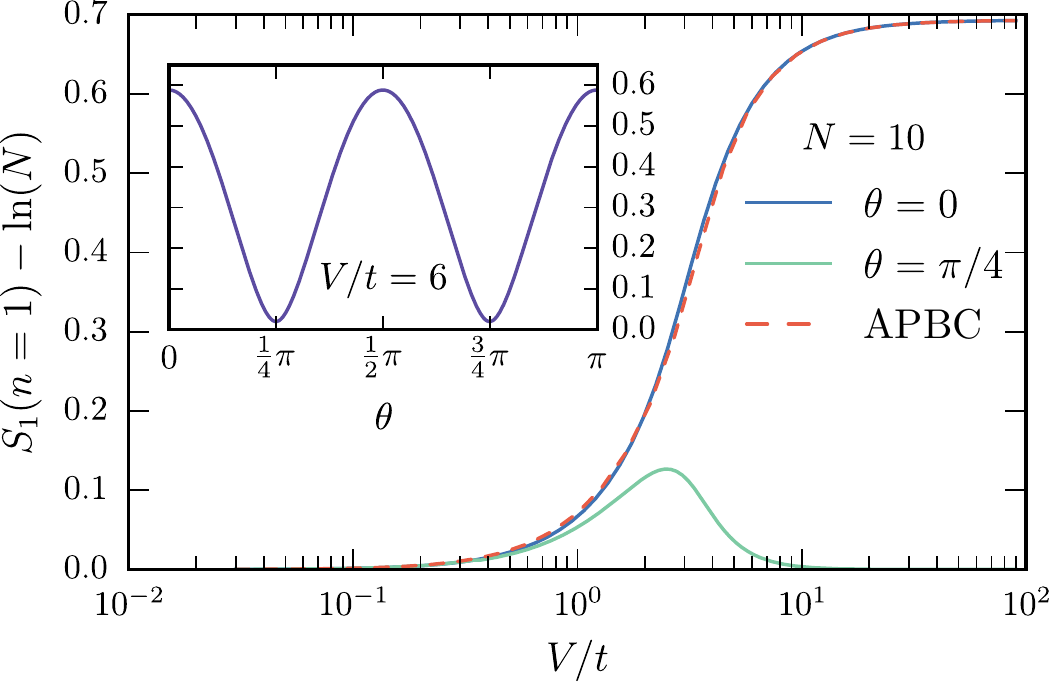}
\end{center}
\caption{Effects of ground state degeneracy. The $S_2(n=1)-\ln(N)$ dependence
on $V/t$ in the ground state of the $t-V$ model for $N = 10$. Solid lines
represent results obtained from the degenerate ground state in
Eq.~(\ref{eq:DGS}) using PBC and $\theta=0,\pi/4$ (see the text for details).
The dashed line corresponds to the non-degenerate ground state for APBC. Inset:
$S_2(n=1)-\ln(N)$ vs $\theta$ for $V/t=6$.}
\label{fig:Sthetadep}
 \end{figure}

Finally we investigate the question of whether particle bipartition EE is
sensitive to the ground state degeneracy known to occur in the $t-V$
model with periodic boundary conditions and an even number of sites.
Introducing the inversion operator $P$ \cite{KampfSekania:2003} defined by
\begin{equation}
Pc^{\dagger}_{i}P^{\dagger}=c^{\dagger}_{M-i+1},
\quad \;i=1, \cdots, M.
\label{eq:inversion}
\end{equation}
where $P$ commutes with the Hamiltonian of the $t-V$ model in
Eq.~(\ref{eq:H-tV}) for PBC, we can write the degenerate ground state as a
superposition of the eigenstates of the inversion operator: $P\vert
\Phi_{\pm}\rangle=\pm\vert \Phi_{\pm}\rangle$, i.e.,
\begin{equation}
\vert \Psi\rangle=\cos(\theta) \vert \Phi_{+}\rangle+\sin(\theta) \vert \Phi_- \rangle. 
\label{eq:DGS}
\end{equation}
Here, we only consider a superposition with real coefficients that can be
varied through the parameter $0\leq\theta\leq\pi$ and study the 
dependence of the \ren EEs on $\theta$ as seen in Fig.~\ref{fig:Sthetadep}. 
%
%
Our numerical results for repulsive interactions with $N=10$ show that
$S_1(n=1)$ oscillates with $\theta$ (Fig.~\ref{fig:Sthetadep} inset), where the
maximum EE corresponds to $\vert \Psi\rangle$ being an eigenstate of $P$, i.e., $\theta=0$ or $\theta=\pi/2$, and the minimum EE is obtained when
both eigenstates $\vert\Phi_{\pm}\rangle$ contribute equally to $\vert
\Psi\rangle$ (maximum uncertainty in $P$, $\theta= \pi/4,3\pi/4$).  Moreover,
the difference between the lower and upper bound vanishes in the non-interacting limit and
widens with increasing interaction strength up to $\ln 2$ in the limit
$V/t\rightarrow \infty$ (see \ref{appendixB}). Interestingly, Fig.~\ref{fig:Sthetadep} shows
that for $\theta=\pi/4$, $S_1(n=1)$ exhibits a peak near the critical
point ($V/t=2$), while the $S_1(n=1)$  dependence on $V/t$ for $\theta=0$ is very
similar to that obtained from the non-degenerate ground state using APBC. 

%
\section{Conclusions}
%

In this paper we have studied the finite size and interaction
dependence of the particle partition \ren entanglement entropies of a fermionic
Tomonaga-Luttinger liquid and find that:
\begin{equation}
    S_\alpha(n,N) = \ln \binom{N}{n} + a_\alpha(n) +
    \mathcal{O}\left(\frac{1}{N^{\gamma_\alpha(n)}}\right)
\label{eq:FinalScaling}
\end{equation}
where $n$ is the number of particles in the subsystem and $\alpha$ the \ren
index.  This result is in agreement with the empirical prediction made in
Ref.~\cite{Zozulya:2007jw}.  For the special case $n=1,\ \alpha=2$ we have determined
the power of the finite size correction to the leading logarithm to be
$\gamma_2(1) = K + K^{-1} - 1$ where $K$ is the Luttinger parameter and
confirmed this interaction dependence for the $t-V$ model by mapping it to the
exactly solvable XXZ chain.  The more general result for $n>1,\ \alpha \ne 2$
in Eq.~(\ref{eq:FinalScaling}) is supported by extensive exact diagonalization
results on the lattice $t-V$ model of spinless fermions obtained on systems
with up to $M=28$ sites.  This general scaling form can be contrasted with a
bosonic Tomonaga-Luttinger liquid, where it was found \cite{Herdman:2015gx}
that $S_2(n,N) \simeq (n/K)\ln N + a'_2(n) + \mathcal{O}(1/N^{1-K^{-1}})$ which
asymptotically recovers the free fermion result in the limit of hard-core
bosons ($K\to 1^+$) using the fact that $\binom{N}{n} \approx N^n/n!$ for $N \gg
n$. 

The universality of the prefactor of the leading order logarithm in
Eq.~(\ref{eq:FinalScaling}) demonstrates that due to the required
anti-symmetrization of the $N$-particle wavefunction, fermions are always more
entangled than bosons under a particle partition. This is consistent with what
was numerically found for hard-core particles with variable anyonic statistics
\cite{Santachiara:2007il}.  Such sensitivity to particle statistics and
interaction dependence is absent in the asymptotic scaling of the spatial mode
entanglement entropy for critical $(1+1)$-dimensional systems where the
prefactor is universal and related to the central charge of the underlying
conformal field theory \cite{Calabrese:2004hl}.  Thus, the particle partition
entanglement appears to be a useful diagnostic of quantum correlations in
many-body systems, and its logarithmic scaling with the total number of
particles $N$ highlights the potential utility of protocols
\cite{Killoran:2014gu} that aim to transfer it to experimentally accessible
mode entanglement.

An interesting open question remains on the origin and development with system
size of the peak in the entanglement entropy in the ground state of
the $t-V$ model near the continuous phase transition at $V/t = 2$ for
macroscopic particle partitions with $n = N/2$
(Fig.~\ref{fig:EntropiesFigure} (b)). A careful finite-size analysis of this
unexpected feature (due to the lack of any natural length scale describing the
partition) would require moving beyond exact diagonalization 
and employing recently adapted hybrid Monte Carlo methods
\cite{Drut:2015fs,Drut:2016el,Porter:2016ft}.

\ack
We thank P.~Fendley and C.~Herdman for enlightening discussions.  The exact
diagonalization code used in this work was adapted from one written by R. Melko
and D.~Iouchtchenko \cite{code}. This research was supported in part by the
National Science Foundation under Award No. DMR-1553991. 

\appendix
\section{Evaluating the $n$-particle partition entanglement}
\label{appendixA}
\renewcommand\thesubsection{\Alph{section}.\arabic{subsection}}

In this appendix, we show that the $n$-RDM of spinless hardcore particles on
a lattice can be written as a tensor product of two lower-rank matrices. This
simplification significantly reduces the numerical cost for calculating $n$-RDM
for such quantum systems. 

In general, for a pure quantum state $\vert \Psi\rangle$ in some
Hilbert space $\cal H$ that can be written as the tensor product space $A \otimes
B$, we can write

\begin{equation}
 \vert \Psi\rangle = \sum_{i,j} C_{i,j} \vert \psi^A_i\rangle  \vert \psi^B_j\rangle
\label{state_decomposition},
\end{equation}
where $\{\vert \psi^A_i\rangle\}$ and $\{\vert \psi^B_j\rangle\}$ are
orthonormal bases in the two Hilbert spaces $A$ and $B$, respectively.
Accordingly, the system degrees of freedom are bipartitioned between the two
subsets  $\{\vert \psi^A_i\rangle\}$ and $\{\vert \psi^B_j\rangle\}$. Using the
product basis $\{\vert \psi^A_i\rangle\vert  \psi^B_j\rangle\}$, the full
density matrix can be written as \begin{equation}
\rho=\vert \Psi\rangle\langle\Psi\vert = \sum_{i,j,i',j'}  \vert \psi^A_i\rangle  \vert \psi^B_j\rangle C_{i,j}C^*_{i',j'} \langle\psi^A_{i'}\vert \langle\psi^B_{j'}\vert 
\label{full_density_matrix}.
\end{equation}
The reduced density matrix $\rho_A$ ($\rho_B$) of subspace $A$ ($B$) , is obtained from $\rho$ by tracing out the degrees of freedom of subspace $B$ ($A$), 
\begin{equation}
\rho_A=\sum_{m}\langle\psi^B_m\vert \rho \vert \psi^B_m\rangle= \sum_{i,j}  \vert \psi^A_i\rangle \left(\sum_{m}C_{i,m}C^*_{j,m}\right) \langle\psi^A_{j}\vert 
\label{rho_1_full},
\end{equation}
\begin{equation}
\rho_B=\sum_{m}\langle\psi^A_m\vert \rho \vert \psi^A_m\rangle= \sum_{i,j}  \vert \psi^B_i\rangle \left(\sum_{m}C_{m,i}C^*_{m,j}\right) \langle\psi^B_{j}\vert 
\label{rho_2_full}.
\end{equation}
 Moreover,  the reduced density matrices can be generated using the linear maps $G_{AB}:S_B\rightarrow S_A$ as  $\rho_A=G_{AB}G_{AB}^\dagger$ and $\rho_B=(G_{AB}^\dagger G_{AB})^T$
where
\begin{equation}
G_{AB}=\sum_{i,j}C_{i,j}\vert \psi^A_i\rangle\langle\psi^B_j\vert 
\label{D_1}.
\end{equation}
Note that, in general, the matrix representing the linear maps $G_{AB}$ is
rectangular since the dimensions of the Hilbert spaces $A$ and $B$ can differ.

\subsection{Particle bipartition}
\label{sec:Particle_bipartition}
Let us now consider a quantum system of $N$ spinless hardcore particles in a
state $\vert \Psi\rangle=\sum_i\chi_i \vert \psi^N_i\rangle$, where $\{\vert
\psi^N_i\rangle\}$ are the $N$ particle second-quantization basis states, where
each basis state corresponds to a single, possible, occupation number
configuration (ONC). Now we recall that each ONC state is a linear combination
of the distinguished particles states $\{\vert \psi^N_{i,j}\rangle\}$ as $\vert
\psi^N_i\rangle=\sum_j\frac{f_j}{\sqrt{N!}}\vert \psi^N_{i,j}\rangle$, where
$j$ runs over all  possible particle permutations (PPs) and $f_j=e^{-i\phi_j}$ is
the corresponding phase factor. Accordingly, we can write 
\begin{equation}
\vert \Psi\rangle=\sum_{i,j}\frac{\chi_if_j}{\sqrt{N!}}\vert \psi^N_{i,j}\rangle.
\label{Psi_pprho_1and2_star}
\end{equation}
%

Now we partition $N$ into two sets of particles: $n_A$ and the remainder 
$n_B=N-n_A$.  The distinguished particles basis $\{\vert \psi^N_{i,j}\rangle\}$
can be written as a tensor product of the two partitions basis 
\begin{equation}
\vert \psi^N_{i,j}\rangle=\vert \psi^{n_A}_{i_A,j_A}\rangle\vert \psi^{n_B}_{i_B,j_B}\rangle
\label{psiN_psin1n2},
\end{equation}
where each ONC (labelled by $i$) of the $N$ particles corresponds to a unique pair of ONCs
$i_A$ and $i_B$ of the $n_A$ and $n_B$ particles, respectively. Similarly, each
PP $j$ of the N particles corresponds to a unique pair of PPs:  $j_A$ and $j_B$
of the $n_A$ and $n_B$ particles.
\begin{equation}
\vert \Psi\rangle=\sum_{i_A,i_B,j_A,j_B}C_{i_A,i_B,j_A,j_B}\vert \psi^{n_A}_{i_A,j_A}\rangle\vert \psi^{n_B}_{i_B,j_B}\rangle
\label{Psi__1and2},
\end{equation}
with
\begin{equation}
C_{i_A,i_B,j_A,j_B}=\frac{\chi_if_j}{\sqrt{N!}}
\label{Ci1i2j1j2}.
\end{equation}
The  $C_{i_A,i_B,j_A,j_B}$ depends on the indices $i$ and $j$ through the
multiplication of $\chi_i$ and $f_j$, and without loss of generality, we can 
take
\begin{equation}
C_{i_A,i_B,j_A,j_B}=\tilde C_{i_A,i_B}\Phi_{j_A,j_B}
\label{Ci1i2Phij1j2}.
\end{equation}
Moreover, the dependence of $\Phi_{j_A,j_B}$ on the PP indices only guarantees that
$\vert \Phi_{j_A,j_B}\vert ^2=constant$ that can be absorbed in $\tilde
C_{i_A,i_B}$. Thus, we can set $\vert \Phi_{j_A,j_B}\vert ^2=1$. Based on the
fact that applying a particle permutation to one group of particles results in an
overall phase factor that does not depend on the permutation of the other group
of particles, we write
\begin{equation}
\Phi_{j_A,j_B}=F^{(A)}_{j_A}F^{(B)}_{j_B}
\label{Phij1j2F1F2},
\end{equation}
with $\vert F^{(A)}_{j_A}\vert ^2=\vert F^{(B)}_{j_B}\vert ^2=1$. Substituting
in Eq.~(\ref{Psi__1and2}) we find
\begin{equation}
\vert \Psi\rangle=\sum_{i_A,i_B,j_A,j_B}\tilde C_{i_A,i_B}F^{(A)}_{j_A}F^{(B)}_{j_B}\vert \psi^{n_A}_{i_A,j_A}\rangle\vert \psi^{n_B}_{i_B,j_B}\rangle
\label{Psi__1and2r},
\end{equation}
Let us now calculate the reduced density matrix of $\rho_A$ using   
\begin{equation}
G_{n_An_B}=\sum_{i_A,i_B,j_A,j_B}\tilde C_{i_A,i_B}F^{(A)}_{j_A}F^{(B)}_{j_B}\vert \psi^{n_A}_{i_A,j_A}\rangle\langle\psi^{n_B}_{i_B,j_B}\vert 
\label{C_1},
\end{equation}
as
\begin{eqnarray}
\rho_A &=& G_{n_An_B}G_{n_An_B}^\dagger\\ &=& \sum_{i^{}_A,j^{}_A,i^{\prime}_A,j^{\prime}_A}\vert \psi^{n_A}_{i^{}_A,j^{}_A}\rangle \sum_{i^{}_B}\left( \tilde C_{i^{}_A,i^{}_B}\tilde C^{*}_{i^{\prime}_A,i^{}_B}\right)F^{(A)}_{j^{}_A}{F}^{*(A)}_{j^{\prime}_A}\sum_{j^{}_B}\left\vert {F}^{(B)}_{j^{}_B}\right\vert ^2\langle\psi^{n_A}_{i^{\prime}_A,j^{\prime}_A}\vert \nonumber\\
 &=&n_B!\sum_{i^{}_A,j^{}_A,i^{\prime}_A,j^{\prime}_A}\vert \psi^{n_A}_{i^{}_A,j^{}_A}\rangle D_{i^{}_A,i^{\prime}_A}\Phi_{j^{}_A,j^{\prime}_A}\langle\psi^{n_A}_{i^{\prime}_A,j^{\prime}_A}\vert 
\label{rho_1_f},
\end{eqnarray}
with $D_{i^{}_A,i^{\prime}_A}= \sum_{i^{}_B} \tilde C_{i^{}_A,i^{}_B}\tilde
C^*_{i^{\prime}_A,i^{}_B}$ and
$\Phi_{j^{}_A,j^{\prime}_A}=F^{(A)}_{j^{}_A}{F}^{*(A)}_{j^{\prime}_A}$.  From
Eq.~(\ref{rho_1_f}) we see that $\rho_A$ is a Kronecker product (tensor
product) of the lower-rank Hermitian matrices $D$ and $\Phi$.  where $D$ can be
calculated considering a single PP for each particle partition and the
elements of $\Phi$ are the product of the relative phases of the chosen
partitions (\ref{Phij1j2F1F2}) 

\subsection{Eigenvalues}
\label{sec:Eigenvalues}

Let $V_D$ and $V_{\Phi}$ be two unitary transformations that diagonalize the
sub matrices $D$ and $\Phi$, respectively. Such that
$V^{\dagger}_DDV^{}_D=\Lambda$ and $V^{\dagger}_{\Phi}\Phi V^{}_{\Phi}=W$,
where $\Lambda$  and $W$ are diagonal matrices with eigenvalues $\{\lambda_k\}$
and $\{w_l\}$.  If we construct the unitary transformation $U$ as
\begin{equation}
U=V_D \otimes V_{\Phi}
\label{U},
\end{equation}
and calculate $U^\dagger(\rho_A/n_B!)U$ we find
\begin{equation}
    U^\dagger\left(\frac{\rho_A}{n_B!}\right)U=\sum_{k,l}\vert \psi^{n_1}_{k,l}\rangle \lambda_k w_l\langle\psi^{n_1}_{k,l}\vert 
\label{UdrhoU}.
\end{equation}
Accordingly, the unitary transformation $U$ diagonalizes $\rho_A$ and the
eigenvalues of $\rho_A$ are $n_B!\lambda_k w_l$. Moreover, $\Phi$ has the
structure of a simple projection operator onto the non-normalized state $\vert
F^{(A)}\rangle=\sum_j^{n_A!} F^{(A)}_j\vert j\rangle=\sum_j^{n_A!}
e^{i\phi_j}\vert j\rangle$ as $\Phi=\vert F^{(A)}\rangle\langle F^{(A)}\vert$.
The only eigenstate of $\Phi$ with a nonzero eigenvalue is $\vert
F^{(A)}\rangle$, where $\Phi\vert F^{(A)}\rangle=\vert F^{(A)}\rangle\langle
F^{(A)}\vert F^{(A)}\rangle=n_A!\vert F^{(A)}\rangle$. 

Therefore, we conclude that the nonzero eigenvalues of $\rho_A$ are
$n_A!n_B!\lambda_k$, where $\lambda_k$ are the eigenvalues of the matrix $D$
that is constructed using only one PP of each of the sets $\{\vert
\psi^{n_A}_{i_A,j_A}\rangle\}$ and $\{\vert \psi^{n_B}_{i_B,j_B}\rangle\}$.
As the rank of $D$ is smaller than that of the $n$-RDM by a factor of
$n_A!n_B!$ the numerical effort involved in calculating the
eigenvalues of the $n$-RDM is enormously reduced.
\section{$n$-particle partition entanglement in the $V/t \to \infty$ limit} 
\label{appendixB}

Here we calculate the $n$-particle partition entanglement of the
one-dimensional fermionic $t-V$ model at half filling ($N=M/2$) in the infinite
repulsion limit ($V/t \rightarrow \infty$). In this limit, the Hamiltonian of
the model (Eq. (\ref{eq:H-tV})) is reduced to
\begin{equation}
  H= V\sum_i n_i n_{i+1}\,
  \label{eq:H-tV_infty}
\end{equation}
which is diagonal in the occupation number representation with a two-fold
degenerate ground state, where, at half filling, the fermions can avoid having
any nearest neighbors by occupying sites with only odd indices
($\vert\psi_{\rm{odd}}\rangle=\vert1010\cdots10\rangle$) or only even indices
($\vert\psi_{\rm{even}}\rangle=\vert0101\cdots01\rangle$). Thus, one can write the
ground state in this limit, as a superposition of
$\vert\psi_{\rm{odd}}\rangle$ and $\vert\psi_{\rm{even}}\rangle$:
\begin{equation}
\vert\Psi\rangle= \cos(\Theta)\e^{i\delta}\vert\psi_{\rm{odd}}\rangle+\sin(\Theta)\vert\psi_{\rm{even}}\rangle,
  \label{eq:gs_infty_deg}
\end{equation}
where we parametrize the amplitudes and the relative phase of the odd/even 
states using $\Theta$ and $\delta$. Note that for $\delta=0$ and
$\Theta=\pi/4$ ($\Theta=3\pi/4$), the ground state $\vert\Psi\rangle$ is also
an eigenstate of the inversion operator $P$ (Eq. (\ref{eq:inversion})) with
eigenvalue $\pm 1$ where
\begin{equation}
P\vert\Phi_{\pm}\rangle=\pm\vert\Phi_{\pm}\rangle =\pm\left(\frac{1}{\sqrt{2}}\vert\psi_{\rm{odd}}\rangle\pm\frac{1}{\sqrt{2}}\vert\psi_{\rm{even}}\rangle\right).
\end{equation}
The degeneracy persists in the case of finite interaction $V/t$  for even/odd
$N$ with PBC/APBC. The degeneracy is lifted for odd/even $N$ with APBC/PBC
with the resulting ground state in the infinite repulsion limit approaching
an eigenstate of $P$:
\begin{equation}
\vert\Psi\rangle=\vert\Phi_+\rangle= \frac{1}{\sqrt{2}}\vert\psi_{\rm{odd}}\rangle+\frac{1}{\sqrt{2}}\vert\psi_{\rm{even}}\rangle.
  \label{eq:gs_infty_NONdeg}
\end{equation}

We now consider the $n$-particle partition entanglement of the degenerate ground
state $\vert\Psi\rangle$ defined in Eq. (\ref{eq:gs_infty_deg}), where we can
write the corresponding full density matrix $\rho$ as 
\begin{eqnarray}
\rho &=\cos^2(\Theta)\vert\psi_{\rm{odd}}\rangle\langle\psi_{\rm{odd}}\vert+\sin^2(\Theta)\vert\psi_{\rm{even}}\rangle\langle\psi_{\rm{even}}\vert\nonumber\\
&\quad+\sin(\Theta)\cos(\Theta)\e^{i\delta}\vert\psi_{\rm{odd}}\rangle\langle\psi_{\rm{even}}\vert+\sin(\Theta)\cos(\Theta)\e^{-i\delta}\vert\psi_{\rm{even}}\rangle\langle\psi_{\rm{odd}}\vert,
\label{eq:gs_rho_deg}
\end{eqnarray}
If we partition the $N$ particles into two distinguishable sets 
of $n_A=n$ and $n_B=N-n$ particles, we can write the states $\vert
\psi_{odd}\rangle$ and $\vert \psi_{even}\rangle$ in terms of the
first-quantized basis states of the two partitions as
\begin{equation}
\vert \psi_{\rm{odd}}\rangle=\sum_{i_A,i_B,j_A,j_B} \frac{f_{i_A,i_B,j_A,j_B}^{\rm{odd}}}{\sqrt{N!}}\vert \psi^{n_A,\rm{odd}}_{i_A,j_A}\rangle\vert \psi^{n_B,\rm{odd}}_{i_B,j_B}\rangle
\label{Psi_odd_AB},
\end{equation}
\begin{equation}
\vert \psi_{\rm{even}}\rangle=\sum_{i_A,i_B,j_A,j_B} \frac{f_{i_A,i_B,j_A,j_B}^{\rm{even}}}{\sqrt{N!}}\vert \psi^{n_A,\rm{even}}_{i_A,j_A}\rangle\vert \psi^{n_B,\rm{even}}_{i_B,j_B}\rangle
\label{Psi_even_AB},
\end{equation}
where the indices $i_A$ and $i_B$ label possible  occupation number
configurations (ONCs) in both partitions $A$ and $B$ while $j_A$ and $j_B$
label different particle permutations (PPs). Also, $f_{i_A,i_B,j_A,j_B}^{\rm{odd}}$ and
$f_{i_A,i_B,j_A,j_B}^{\rm{even}}$ are overall phase factors, where the
superscript odd (even) is to indicate that only sites with odd (even) indices
are occupied.  We note that in this decomposition  the states
$\vert\psi_{\rm{even}}\rangle$ and $\vert \psi_{\rm{odd}}\rangle$ 
are constructed from non-overlapping subspaces (even/odd) of partition $B$.
Similarly for partition $A$.
By tracing out all degrees of freedom in $B$ from $\rho$ (Eq.
(\ref{eq:gs_rho_deg})), we can write the reduced density matrix $\rho_A$ as
\begin{equation}
    \rho_A = {\Tr}_{B}\,\rho= \cos^2(\Theta){\Tr}_{B}\,\vert\psi_{\rm{odd}}\rangle\langle\psi_{\rm{odd}}\vert+\sin^2(\Theta) {\Tr}_{B}\, \vert\psi_{\rm{even}}\rangle\langle\psi_{\rm{even}}\vert,
\label{eq:rho_A_final}
\end{equation}
where the trace of the mixed terms
($\vert\psi_{\rm{odd}}\rangle\langle\psi_{\rm{even}}\vert$,
$\vert\psi_{\rm{even}}\rangle\langle\psi_{\rm{odd}}\vert$) vanishes due to the
non-sharing of $B$ basis states.  Moreover, 
$\rho_A^{\rm odd}={\Tr}_{B}\,\vert\psi_{\rm{odd}}\rangle\langle\psi_{\rm{odd}}\vert$
and $\rho_A^{\rm even}={\Tr}_{B}\,
\vert\psi_{\rm{even}}\rangle\langle\psi_{\rm{even}}\vert$ contribute separately
to the spectrum of $\rho_A$ due to the non-sharing of $A$ basis states.

We now calculate the spectrum of $\rho_A^{\rm odd}$. Note that the state $\vert
\psi_{\rm odd}\rangle$ represents a single ONC of the $N$ particles and as a
result the ONC $i_A$ is uniquely determined by $i_B$ in the product states
$\vert \psi^{n_A,\rm{odd}}_{i_A,j_A}\rangle\vert
\psi^{n_B,\rm{odd}}_{i_B,j_B}\rangle$. Therefore, $\rho_A^{\rm odd}$ does not
connect any pair of states, in the set $\{\vert
\psi^{n_A,\rm{odd}}_{i_A,j_A}\rangle\}$, with different ONC $i_A$. This result,
combined with the formalism presented in \ref{appendixA}, allows us to identify
that the sector
of $\rho_A^{\rm odd}$ that connects states in $\{\vert
\psi^{n_A,\rm{odd}}_{i_A,j_A}\rangle\}$ with
fixed PP $j_A$ is diagonal with $\binom{N}{n}$ equal non-zero elements of value
$\frac{1}{N!}$.  $\binom{N}{n}$ is the number of possible ONCs in the
partition $A$ with $n_A=n$ and we only consider the contribution of a single PP
$j_B$ to ${\Tr}_{B}\,\vert\psi_{\rm{odd}}\rangle\langle\psi_{\rm{odd}}\vert$.
It then follows from \ref{appendixA} that the non-zero eigenvalues of
$\rho_A^{\rm odd}$  can be obtained by rescaling the above eigenvalues by a factor
of $n_A!n_B!=n!(N-n)!$.  By an equivalent set of arguments 
$\rho_A^{\rm even}$ has the same eigenvalues. Combining all the above and using
Eq.~(\ref{eq:rho_A_final}), we find that $\rho_A$ has two sets of eigenvalues:
$\binom{N}{n}$ eigenvalues of $\cos^2(\Theta)/{\binom{N}{n}}$ and
$\binom{N}{n}$ eigenvalues of $\sin^2(\Theta)/{\binom{N}{n}}$. Therefore, 
the \ren entanglement entropies are
\begin{equation}
S_{\alpha}(n) = \ln
\binom{N}{n}+
\frac{1}{1-\alpha} \ln\left[\cos^{2\alpha}(\Theta)+\sin^{2\alpha}(\Theta)\right]
\label{eq:S_alpha_deg},
\end{equation}
and the von Neumann entropy ($\alpha = 1$) is
\begin{equation}
S_1(n) = \ln \binom{N}{n}-\cos^2(\Theta)
\ln\left[\cos^2(\Theta)\right]-\sin^2(\Theta)\ln\left[\sin^2(\Theta)\right].
\label{eq:S1_deg}
\end{equation}
According to Eqs.~(\ref{eq:S_alpha_deg}) and (\ref{eq:S1_deg}), the maximum
entropy corresponds to $\Theta=\pi/4$ and $3\pi/4$ ($\vert\Psi\rangle=
\frac{\e^{i\delta}}{\sqrt{2}}\vert\psi_{\rm{odd}}\rangle+\frac{1}{\sqrt{2}}\vert\psi_{\rm{even}}\rangle$),
where all the $2\binom{N}{n}$ eigenvalues of $\rho_A$ are equal and thus all
the \ren entropies are equal to
\begin{equation}
S_{\alpha}(n) = \ln \binom{N}{n}+\ln2.
\label{eq:rho_A_final1}
\end{equation}
For $\Theta=0$ and $\pi/2$,  $\vert\Psi\rangle= \vert\psi_{\rm{odd}}\rangle$ or
$\vert\psi_{\rm{even}}\rangle$, only $\binom{N}{n}$ equal eigenvalues survive
yielding a minimum entropy of
\begin{equation}
S_{\alpha}(n) = \ln \binom{N}{n}.
\label{eq:rho_A_final2}
\end{equation}
These limits can be seen in Fig.~\ref{fig:Sthetadep} for $V/t \gg 1$.

 \section*{References}
\bibliography{refs}

\end{document}